# The KdV hierarchy in terms of whole powers of an integro-differential operator

B. P. Ryssev

**Abstract**
It is shown that equations of the Korteweg-de Vries hierarchy and their conservation laws can be expressed via the whole powers of an integro-differential operator and functions provided by them.

**Keywords**: KdV hierarchy, conservation laws, powers of operator.     MSC: 37K10, 35L65

**1.** The KdV hierarchy (KdVH) is the infinite family of nonlinear partial differential equations of evolution type for a real valued function $u(x, t)$, $m$ is odd

$$U_m \equiv u_t + K_m(u, u_x, \ldots, u^{(m)}) = 0 \qquad K_m = (J_m)_x$$

(for a function $f(x)$, $f^{(i)} = \partial^i f/\partial x^i$ in general and for $i > 3$, the x-derivatives for $i \leq 3$ are denoted by subscript). Successive fluxes $J_m$ are generated by the recursion operator $R$

$$J_{m+2} = RJ_m \qquad R = \partial^2 + 2u + 2\partial^{-1}u\partial \qquad \partial^i = \partial^i/\partial x^i,\ \partial\partial^{-1} = \partial^{-1}\partial = 1$$

starting from $J_1 = u$. The next two are ($m = 3$ corresponds to the KdV equation)

$$J_3 = Ru = u_{xx} + 3u^2 \qquad J_5 = RJ_3 = u^{(4)} + 5(2uu_{xx} + u_x^2 + 2u^3)$$

Each equation $U_m = 0$ possesses infinite number ($k$) of conservation laws

$$(\rho_k)_t + (J_{km})_x = 0$$

The first is the equation itself with $\rho_1 = u$. The left-hand sides of all of them can be expressed as homogeneous functions of $U_m$ of the first degree (§4), and therefore hold on solutions of $U_m = 0$. Each density $\rho_k$ is common for the entire hierarchy (we mark them by $k$ odd, see §2 for the reason).

Further we widely use the notions of the Frechet derivative and Euler operator [1]. For a differential function $\psi(u, u_x, \ldots, u^{(n)})$, its Frechet derivative in the direction of $f$, $\psi'[f]$, is defined as

$$\psi'[f] = d\psi(u + \varepsilon f)/d\varepsilon \,|_{\varepsilon = 0} \qquad (1)$$

It maps $\psi$ to the linear differential operator, $i = 0, \ldots, n$

$$\psi' = \Sigma(\partial\psi/\partial u^{(i)})\partial^i \qquad (2)$$

It has properties $(\psi_1\psi_2)' = \psi_1\psi_2' + \psi_2\psi_1'$ and $(\psi_x)' = \partial\psi'$. {Let $\psi = u^2/2$, then $(\psi_x)' = (uu_x)' = u_x + u\partial = \partial u = \partial\psi'$}. (Inside the curly brackets, like here, we put sometimes short illustrative examples).

The Euler operator, $i = 0, \ldots, n$

$$E = \Sigma(-\partial)^i(\partial/\partial u^{(i)})$$

has the property ($O^\dagger$ is adjoint of an operator $O$)

$$E\psi_1\psi_2 = \psi_1'^\dagger[\psi_2] + \psi_2'^\dagger[\psi_1] \qquad \text{hence} \qquad E\psi = \psi'^\dagger[1] \qquad (3a, b)$$

We say that $\psi$ is the exact derivative (ED), if it can be represented as the x-derivative of another function. {$\psi \equiv 2uu_{xxx} = (2uu_{xx} - u_x^2)_x$, so $\psi$ is ED}.

If two operators $O_1$ and $O_2$ are related as $O_1 = \partial O_2$, we say that $O_1$ is $\partial O$. {$(\psi_x)'$ is $\partial O$}. With this notion, the definition of adjoint of an operator $A$ can be written in operator form: $fA - (A^\dagger f) = \partial O$.

Since $(\partial^i)^\dagger = (-\partial)^i$, then from definitions of $\psi'$ and $E$ we have

$$\psi' - E\psi = \partial O \qquad (4)$$

where the function $E\psi$ serves as a multiplication operator. {$J_3' = \partial^2 + 6u$, $EJ_3 = 6u$}.

By formula $(O_1O_2)^\dagger = O_2^\dagger O_1^\dagger$ we have $(\partial O)^\dagger = -O^\dagger\partial$, hence $(\partial O)^\dagger 1 = 0$. The use of this in the adjoint of (4), $\psi'^\dagger - E\psi = (\partial O)^\dagger$, justifies (3b); which in turn gives $E\psi_x = (\partial\psi')^\dagger[1] = 0$. With that, if $\psi = \psi_1 + $ ED, then $E\psi = E\psi_1$. Such two functions are called equivalent.

We also deal with the functions involving monomials $p \equiv (\partial^{-1}u)$, $p^2(\partial^{-1}u^2)$ and the like. The Frechet derivative of such, let us call it so, nonlocal function $\Psi$ can be evaluated using property $(\partial^{-1}\psi)' = \partial^{-1}\psi'$, formula (2) with inclusion of $i = -1$, and directly (1).

$\{p' = \partial^{-1}$; $(p^n)' = (\partial p^n/\partial p)\partial^{-1}$; $(up)' = up' + pu' = u\partial^{-1} + p = \partial p \partial^{-1}$ is $\partial O$ as $up = (p^2)_x/2$ is ED$\}$.

Using the fact that $\partial^{-1}$ is skew-adjoint, $f\partial^{-1} + (\partial^{-1}f) = \partial(\partial^{-1}f)\partial^{-1}$, we can apply (3) to $\Psi$.

$\{p'^{\dagger} = -\partial^{-1}$, then by (3a): $Eup = u'^{\dagger}[p] + p'^{\dagger}[u] = p - (\partial^{-1}u) = 0\}$.

One more property of $\psi'$. Let $u_y \equiv \partial u/\partial y$, where $y$ is $x$ or $t$, then $\psi'[u_y] = \psi_y$. The same for $\Psi$.

$\{(u_x^2)'[u_y] = 2u_x(u_y)_x = (u_x^2)_y$; $p'[u_y] = (\partial^{-1}u_y) = p_y\}$.

Conventions. While $(O\psi)$ always denotes a function, $O\psi$ can stand either for a function (like $RJ_m$ and $E\psi$) or an operator, the meaning is usually clear from the context. (The notation of such an operator using the composition sign, $O \circ \psi$, does not help fully avoid the ambiguity, it only makes many formulas difficult to read). Indices $m$ and $k$ are odd throughout; $c$, $c_n$ are constants; l.h.s and r.h.s stand for left and right-hand-side.

**2.** Our representation of the KdVH employs the integro-differential operator [2]

$$X = \partial + u\partial^{-1}$$

its whole powers $X^n$ and their action on $u$: the functions $f_{n+1} = X^n u$, $f_1 = u$. For comparison, the well known presentation of the KdVH uses the operator $L = \partial^2 + u$ and its fractional powers $L^{m/2}$ [1]. Unlike $X^n$ and $X^n u$, $L^{m/2}$ is an infinite series with respect to both $\partial^n$ and $\partial^{-n}$, and $L^{m/2}u$ is *not defined* in principle.

We begin with these $f_n$ related as

$$f_{n+1} = Xf_n \equiv (f_n)_x + u\varphi_n \qquad \varphi_n \equiv (\partial^{-1}f_n)$$

The first few of them are (integration constants are assumed to be zero everywhere), $p \equiv \varphi_1$

$$f_2 = Xu = u_x + up = [u + (1/2)p^2]_x \tag{5}$$

$f_3 = (f_2)_x + u\varphi_2 = u^2 + (y_3)_x \qquad y_3 = f_2 + (1/6)p^3$

$f_4 = [f_3 + (1/2)u^2 + p(\partial^{-1}u^2) + (1/4!)p^4]_x$

$f_5 = 2u^3 + uu_{xx} + (y_5)_x \qquad y_5 = f_4 + (1/2)pu^2 + (1/2)p^2(\partial^{-1}u^2) + (1/5!)p^5$

$f_7 = 5u^4 + uu^{(4)} + 8u^2u_{xx} + 6uu_x^2 + (y_7)_x$

$y_7 = f_6 + [(1/2)u^2 + (1/24)p^4](\partial^{-1}u^2) + 2pu^3 + (1/2)p^2(\partial^{-1}uu_{xx}) + p^2(\partial^{-1}u^3) + (1/2)p(\partial^{-1}u^2)^2 + (1/12)u^2p^3 + (1/2)p(u^2)_{xx} - (3/2)pu_x^2 + (1/7!)p^7$

**Statement 1**. All $f_n$ with $n$ even are EDs.

Proof. Introduce the generating function of $f_n$

$$\Phi = \Sigma f_n z^n = \Phi_o + \Phi_e$$

where $\Phi_o$ and $\Phi_e$ are series with odd and even $n$. They are related as, $\zeta \equiv 1/z$,

$$X\Phi_o = \zeta\Phi_e \qquad X\Phi_e = \zeta\Phi_o - u \tag{6a, b}$$

Multiply (6a, b) by $\varphi_{e, o} \equiv (\partial^{-1}\Phi_{e, o})$ respectively, subtract one product from another and take $u\varphi_o$ from (6a) to get

$$\Phi_e = (zJ + \delta)_x \qquad J \equiv \Phi_o + \Phi_o\varphi_e - \Phi_e\varphi_o \qquad \delta \equiv (1/2)(\varphi_o^2 - \varphi_e^2) \tag{7}$$

Thus, $\Phi_e$, and hence each its member, is ED.

$\{\Phi_e(z^4) = [J(z^3) + \varphi_1\varphi_3 - (1/2)\varphi_2^2]_x$, $J(z^3) = f_3 + f_1\varphi_2 - f_2\varphi_1$. This $\Phi_e(z^4)$ is another form of $f_4\}$.

**Statement 2**. The series $J$ in (7) is such that $J(z^m) = J_m$, meaning $J$ is the generating function of fluxes $J_m$, that is, $J = \Sigma J_m z^m$. Consequently, the series $K \equiv J_x$ is that of $K_m$.

The second form of $J$, by (7), is $J = \zeta(\varphi_e - \delta)$. Hence $K = \zeta(\Phi_e - \Phi_o\varphi_o + \Phi_e\varphi_e)$.

First, check some equalities $J(z^m) = J_m$ by inspection: $J(z^1) = f_1 = J_1$ and

$$J(z^3) = (f_2)_x + 2u\varphi_2 - f_2 p = (u_x + pu)_x + 2u(u + p^2/2) - (u_x + pu)p = u_{xx} + 3u^2 = J_3$$

In $J(z^5)$ too, all nonlocal components cancel leaving $J_5$.

Proof. We are to show that the adjacent members of J are connected by the recursion operator R. This operator naturally appears in $X\Phi_e$. With $\Phi_e$ from (7)

$$X\Phi_e \equiv (\Phi_e)_x + u\varphi_e = (zJ + \delta)_{xx} + u(zJ + \delta) = z(J_{xx} + uJ) + B$$

$B \equiv \delta_{xx} + u\delta = \varphi_o(\Phi_o)_x - \varphi_e(\Phi_e)_x + \Phi_o^2 - \Phi_e^2 + u\delta$. Here

$$\Phi_o^2 - \Phi_e^2 = z(\partial^{-1} 2uK) \tag{8}$$

which follows from the difference of (6a) multiplied by $2\Phi_o$ and (6b) by $2\Phi_e$. The other terms in B, with $(\Phi_{e,o})_x$ taken from (6), become

$$\varphi_o(\zeta\Phi_e - u\varphi_o) - \varphi_e(\zeta\Phi_o - u\varphi_e - u) + u\delta = \zeta(\varphi_o\Phi_e - \varphi_e\Phi_o) + u(\varphi_e - \delta) = \zeta(\Phi_o - J) + zuJ$$

So, $X\Phi_e = z[J_{xx} + 2uJ + (2\partial^{-1}uK)] + \zeta(\Phi_o - J)$, where [.] = RJ. This, due to (6b), becomes

$$RJ = \zeta^2(J - zJ_1) \tag{9}$$

That is, $RJ(z^m) = J(z^{m+2})$. The claim is proved.

**Definition**: The differential function $\rho_{kD}$ in $f_k = \rho_{kD} + (y_k)_x$, k odd, is the density common for the entire hierarchy. (See §4 for the proof; subscript D means definition). From (5)

$$\rho_{3D} = u^2 \qquad \rho_{5D} = 2u^3 + uu_{xx} \qquad \rho_{7D} = 5u^4 + uu^{(4)} + 8u^2 u_{xx} + 6uu_x^2$$

Each $f_k$ contains at least one monomial $u^\alpha$, $\alpha = (k + 1)/2$, and thus cannot be ED as a whole. Indeed, $f_k = X^2 f_{k-2}$, where $X^2 = \partial^2 + 2u + u_x\partial^{-1} + u\partial^{-1}u\partial^{-1}$ involves $2u$. So, starting from $f_1 = u$, the power of u increases by one, step by step, thereby producing such $\alpha$. In fact, as is seen, $\rho_k$ contains the term $c_k u^\alpha$; these $c_k$ can be found by formula (16). Also, $\partial^2$ in $X^2$ gives rise to $uu^{(k-3)}$ in $f_k$.

Some comments on the above $\rho_{kD}$'s:
- By construction, $(y_k)_x = (f_{k-1})_x + \Psi_x$, so $u\varphi_{k-1} = \rho_{kD} + \Psi_x$, $\Psi$ is a nonlocal function.
- As is seen, $\rho_{kD}$ is a weighted sum of all monomials from $uJ_{k-2}$. Indeed, due to (7), $u\varphi_e = zuJ + u\delta$ (that is some of them arise from $u\delta$). $\{\rho_{5D} = uJ_3 - u^3\}$. Even though the structure of $\rho_{kD}$ is known, retrieving $\rho_{kD}$ from $f_k$ is still an ad hoc procedure.
- $E\rho_{kD} = 2J_{k-2}$. $\{E\rho_{5D} = 6u^2 + 2u_{xx} = 2J_3\}$. Hence $Ef_k = Eu\varphi_{k-1} = 2J_{k-2}$; cf. (13).
- $\rho_{7D}$ contains an implicit ED: $\rho_{7D} = 5u^4 + uu^{(4)} + 5u^2 u_{xx} + (u^3)_{xx}$. (Remark, one can include such ED in $(y_k)_x$ and redefine $\rho_{kD}$ as a nonderivative part of $f_k$).

For illustrations and comparison with our results we use some densities $\rho_k$ from [3] (found there for the KdV in the form $u_t + u_{xxx} + uu_x = 0$, they are appropriately rescaled here and indexed according to our scheme)

$$\rho_3 = \rho_{3D} = u^2 \qquad \rho_5 = 2u^3 - u_x^2 \qquad \rho_7 = 5u^4 + u_{xx}^2 - 10uu_x^2$$
$$\rho_9 = 14u^5 - u_{xxx}^2 + 14uu_{xx}^2 - 70u^2 u_x^2$$

These and each of ten densities in [3] involves the quadratic term $(-1)^i (u^{(i)})^2$, $i = (k - 3)/2$, $k > 3$. It is equivalent to $uu^{(k-3)}$ in $\rho_{kD}$ and can be obtained as the final integrand of successive integrations of $uu^{(k-3)}$ by parts (which in the end yields the conserved quantities, constants of motion). All $\rho_k$ satisfy $E\rho_k = 2J_{k-2}$. Some equivalency relations: $\rho_{5D} = \rho_5 + (1/2)(u^2)_{xx}$, $\rho_{7D} = \rho_7 + (uu_{xx} - u_x^2 + (8/3)u^3)_{xx}$.

**3.** Here we prove some properties of the constituents of the KdVH. The generating functions introduced above help do this.
- $K_k J_m$ = ED for any k and m (both odd). (10)

Proof. The formula

$$f_x Rf \equiv f_x[f_{xx} + 2uf + 2(\partial^{-1}uf_x)] = f_x f_{xx} + 2[f(\partial^{-1}uf_x)]_x$$

shows that $f_x Rf$ = ED. Let $f = R^n J$, then using (9) (with z = 1 for readability), $RJ = J - J_1$, we have

$$(R^n J)_x R(R^n J) = (R^n J)_x R^n(RJ) = (R^n J)_x R^n(J - J_1) = f_x f - f_x R^n J_1 = ED$$

Hence the products $(R^n J)_x R^n J_1 = ED$. For n = 0, 1, 2 they are

$$KJ_1 \qquad (K - K_1)J_3 \qquad (K - K_1 - K_3)J_5$$

The last is due to $R^2 J = R(J - J_1) = J - J_1 - J_3$. This equals $J_5 + J_7 + \ldots$ In general, each product is $(K_m + K_{m+2} + \ldots)J_m = ED$. Hence $K_{m+j}J_m = ED$, j even. Then $J_{m+j}K_m = ED$ too, as the sum of these two pairs is ED. This completes the proof.

$$\{2K_1 J_5 \equiv 2u_x[u^{(4)} + 5(2uu_{xx} + u_x^2 + 2u^3)] = (2u_x u_{xxx} - u_{xx}^2 + 10uu_x^2 + 5u^4)_x\}.$$

- $J_m'$ are self-adjoint operators.

Proof. Since $K_k J_m = ED$, by (10), then $0 = EK_k J_m = (\partial J_k')^\dagger[J_m] + J_m'^\dagger[K_k]$, by (3a). The first term is $-J_k'^\dagger \partial[J_m] = -J_k'^\dagger[K_m]$. So, $J_m'^\dagger[K_k] = J_k'^\dagger[K_m]$. For k = 1: $J_m'^\dagger[u_x] = K_m$, where $K_m = J_m'[u_x]$, hence $J_m'^\dagger = J_m'$. The claim is proven, and we can drop the dagger in the general case to get

$$J_m'[K_k] = J_k'[K_m] \tag{11}$$

- $EJ_k = 2kJ_{k-2}$ (12)

Proof. For $RJ \equiv J_{xx} + 4uJ - 2(\partial^{-1} u_x J)$ we have $(RJ)' = RJ' + 4J - 2\partial^{-1}J\partial$. Then $(RJ)'[1] = REJ + 4J$, by (3b), as $J'^\dagger = J'$. On the other hand, $(RJ)' = \zeta^2 J' - \zeta$, by (9), so $(RJ)'[1] = E(RJ + \zeta u) - \zeta = ERJ$. Equating these two forms yields

$$ERJ = REJ + 4J \qquad \text{that is} \qquad EJ_k = REJ_{k-2} + 4J_{k-2}$$

If the claim is true, then the r.h.s is $2(k-2)RJ_{k-4} + 4J_{k-2} = 2kJ_{k-2}$. This induction completes the proof. (Note, $EJ_k = J_k'[1] = \partial J_k/\partial u \equiv (J_k)_u$, by (3b) and the definition (2) of $\psi'$).

$\{(J_3)_u \equiv (u_{xx} + 3u^2)_u = 6J_1, (J_5)_u \equiv [u^{(4)} + 5(2uu_{xx} + u_x^2 + 2u^3)]_u = 10J_3$ and $(J_7)_u = 14J_5$, where $J_7 = RJ_5 = u^{(6)} + 35u^4 + 7(2uu^{(4)} + 3u_{xx}^2 + 4u_x u_{xxx}) + 70(u^2 u_{xx} + uu_x^2)\}.$

- $E\Phi_o = 2z^2 J + z$ that is $Ef_k = Eu\varphi_{k-1} = 2J_{k-2}$ (13)

Proof. First we find $\delta'^\dagger[u]$. From $\delta' = \varphi_o \varphi_o' - \varphi_e \varphi_e'$ we have $\delta'^\dagger = \varphi_o'^\dagger \varphi_o - \varphi_e'^\dagger \varphi_e$ and thus $\delta'^\dagger[u] = \varphi_o'^\dagger[\varphi_o u] - \varphi_e'^\dagger[\varphi_e u]$ (operators $\varphi_{o,e}'^\dagger$ act on functions). With $u\varphi_{o,e}$ from (6), this becomes

$$\delta'^\dagger[u] = \varphi_o'^\dagger[\zeta\Phi_e - (\Phi_o)_x] - \varphi_e'^\dagger[\zeta\Phi_o - (\Phi_e)_x - u]$$

The use of $\varphi'^\dagger = -\Phi'^\dagger \partial^{-1}$ (as $\varphi' = \partial^{-1}\Phi'$) in the first, second and fourth terms gives

$$\delta'^\dagger[u] = -\zeta\Phi_o'^\dagger[\varphi_e] + \Phi_o'^\dagger[\Phi_o] - \zeta\Phi_e'^\dagger[\Phi_o] - \Phi_e'^\dagger[\Phi_e] + \varphi_e'^\dagger[u]$$
$$= -\zeta E\Phi_o \varphi_e + (1/2)E(\Phi_o^2 - \Phi_e^2) + \varphi_e'^\dagger[u] = -EZ + \varphi_e'^\dagger[u], \quad Z \equiv \zeta\Phi_o\varphi_e - (1/2)(\Phi_o^2 - \Phi_e^2)$$

Since $\varphi_e'^\dagger[u] = zJ'^\dagger[u] + \delta'^\dagger[u]$, by (7), we arrive at the equation $EZ - zJ'^\dagger[u] = 0$.
The use of $\Phi_o^2 - \Phi_e^2 = z(\partial^{-1}2uK)$, see (8), and J in the form $J = \Phi_o + 2\Phi_o\varphi_e - (\varphi_o\varphi_e)_x$ gives $2\zeta Z = \zeta^2(J - \Phi_o) - (\partial^{-1}2uK) + ED$. Then (9), $RJ = \zeta^2(J - zJ_1)$, leads to $2\zeta Z = 2uJ + \zeta u - \zeta^2\Phi_o + ED$. As a result, the above equation with EZ (times $2\zeta$) reads

$$E(2uJ + \zeta u - \zeta^2 \Phi_o) - 2J'^\dagger[u] = 2J + \zeta - \zeta^2 E\Phi_o = 0$$

proving the claim.

Since $f_k = \rho_{kD} + (y_k)_x$, then (13) leads to

$$E\rho_{kD} = 2J_{k-2} \tag{14}$$

which justifies and generalizes the property of $\rho_{kD}$ (and their equivalents) noticed earlier. Such relations are known (albeit not thus derived) and underlie the Hamiltonian formalism [1] (with E understood as variational derivative of the conserved quantities, functionals).

Here are some examples of how to deal with individual $f_n'$, using the properties of nonlocal functions, outlined in §1. From (5)

$f_3 = u^2 + (y_3)_x$, $y_3 = f_2 + (1/6)p^3$, $f_2 = u_x + pu$, $(y_3)_x = u_{xx} + u^2 + pu_x + up^2/2$, $\varphi_2 = u + p^2/2$. Then

$(y_3)_x' = \partial^2 + 2u + p\partial + u_x\partial^{-1} + p^2/2 + up\partial^{-1} = \partial^2 + u + p\partial + \varphi_2 + f_2\partial^{-1} = \partial^2 + \partial p + \partial\varphi_2\partial^{-1}$, which reflects the general rule, as the last expression is $\partial y_3'$. Indeed, $y_3' = f_2' + (1/2)p^2\partial^{-1}$, where $f_2' = \partial\varphi_2'$, $\varphi_2' = 1 + p\partial^{-1}$. So, $y_3' = \partial(1 + p\partial^{-1}) + (1/2)p^2\partial^{-1} = \partial + p + \varphi_2\partial^{-1}$.

- $J_k = k\rho_k + \text{ED}$ \hfill (15)

This follows from (12) and (14). Since $J_k = RJ_{k-2} \equiv (J_{k-2})_{xx} + G_k$, $G_k = 2uJ_{k-2} + 2(\partial^{-1}uK_{k-2})$, it means that $G_k = k\rho_k + \text{ED}$ (and $EG_k = 2kJ_{k-2}$). Here are some examples of how $G_k$ can be interpreted.

$G_3 = 3u^2$, hence $J_3 = 3\rho_3 + u_{xx}$

$G_5 = 10u^3 + 4uu_{xx} - u_x^2 = 5\rho_{5D} - (uu_x)_x = 5\rho_5 + 2(u^2)_{xx}$

Hence, via $\rho_5$: $J_5 = 5\rho_5 + (u_{xx} + 5u^2)_{xx}$. Also, the form (15) can be revealed by integration by parts

$(\partial^{-1}G_7) = 7(\partial^{-1}\rho_7) + 4uu_{xxx} - 6u_xu_{xx} + 40u^2u_x$

$(\partial^{-1}G_9) = 9(\partial^{-1}\rho_9) + 4uu^{(5)} - 6u_xu^{(4)} + 8u_{xx}u_{xxx} + 56u^2u_{xxx} - 28uu_xu_{xx} + 280u^3u_x$

- From (14) and (15): $E\rho_{k+2} = 2J_k = 2k\rho_k + \text{ED}$, where $E\rho_{k+2} = (\rho_{k+2})_u + \text{ED}$, by definition of E. So,

$(\rho_{k+2})_u = 2k\rho_k + \text{ED}$

Owing to this, the terms $c_ku^\alpha$ in $\rho_k$ and $c_{k+2}u^{\alpha+1}$ in $\rho_{k+2}$ are related as $(c_{k+2}u^{\alpha+1})_u = 2k(c_ku^\alpha)$. It gives $(\alpha + 1)c_{k+2} = 2kc_k$, where $\alpha = (k + 1)/2$, and thereby the recurrence

$c_{k+2} = [4k/(k + 3)]c_k$ \hfill (16)

Starting from $c_1 = c_3 = 1$ ($\rho_3 = u^2$), this yields $c_5 = 2$, $c_7 = 5$, $c_9 = 14$, etc.. Alternatively, with $J_k$ known, $c_k$ can be found from $J_k/k$, due to (15).

**4**. With the above properties proven, we are ready to derive the conservation laws of the KdVH. For these evolution equations, a function $g(u)$ is a density, if $g'[K_m] = \text{ED}$. Because then the equation $g_t + g'[K_m] = 0$ is the conservation law, and it holds when $U_m = 0$ ($g'[K_m]$ becomes $- g'[u_t] = - g_t$). Or, in the other words, because its l.h.s is $g'[U_m]$. This is the case for $g(u)$ such that $Eg = cJ_k$. Indeed, using (4), $g' = Eg + \partial O$, we have $g'[K_m] = cJ_kK_m + (OK_m)_x = \text{ED}$, as $J_kK_m = \text{ED}$ due to (10). If $g(u)$ contains a derivative, $y_x$, then due to $(y_x)' = \partial y'$, $g'[K_m]$ involves $(y'[K_m])_x$. This together with $(y_x)_t$ forms $(y'[U_m])_x$, the separate trivial part of conservation law.

Owing to (14), (13) and (12), the functions that can serve as densities are $\rho_{kD}$, $f_k$ and $J_k$. They produce the conservation laws

$(\rho_{kD})_t + \rho_{kD}'[K_m] = 0$ \quad $(f_k)_t + f_k'[K_m] = 0$ \quad $(J_k)_t + J_k'[K_m] = 0$ \hfill (17a, b, c)

where (a) is a nontrivial part of (b) and (c), by (15). The first equation justifies the definition of $\rho_{kD}$ as densities and shows that the fluxes $J_{km}$ paired with them are given by $(J_{km})_x = \rho_{kD}'[K_m]$. There is still a trivial part in (17a), when $\rho_{kD}$ has an implicit ED or it is expressed via its equivalent. (Note, if we are interested in conserved quantity only, there is no need even to know $\rho_{kD}'[K_m]$ explicitly).

The equations (17c) represent the infinity of polynomial conservation laws of each equation of the KdVH in the most explicit form. The nontrivial part of each of them, involving an a priori unknown density (times k), can be revealed by integration of $J_k$, as it is done under (15).

The known fact that each $K_k$ is symmetry of equation $U_m \equiv u_t + K_m = 0$ [1], which means

$(K_k)_t + K_m'[K_k] = \{(J_k)_t + J_m'[K_k]\}_x = 0$,

follows immediately from the conservation laws (17c), owing to $J_m'[K_k] = J_k'[K_m]$, see (11).

Examples of the relation (4), $\psi' - E\psi = \partial O$, for $\psi = \rho_5 = 2u^3 - u_x^2$ and $\rho_7 = 5u^4 + u_{xx}^2 - 10uu_x^2$

$\rho_5' = 6u^2 - 2u_x\partial$, $\rho_5' - 2J_3 = -2\partial u_x$

$\rho_7' = 20u^3 + 2u_{xx}\partial^2 - 10u_x^2 - 20uu_x\partial$, $\rho_7' - 2J_5 = 2\partial(u_{xx}\partial - u_{xxx}) - 20\partial uu_x$

Since $(y_x)' = \partial y'$, (4) holds for $\rho_{kD}$ as well. Two examples of $(J_{k3})_x$

$\rho_3'[K_3] = 2uK_3 = (4u^3 + 2uu_{xx} - u_x^2)_x$ \qquad $\rho_5'[K_3] = 2(J_3 - \partial u_x)K_3 = (J_3^2 - 2u_xK_3)_x$

One examples of (4) for $\psi = J_5$ using $EJ_k = J_k'[1]$ (note, $J_k' - J_k'[1] = \partial O$ implies that $\partial O = \partial O_1 \partial$)
$$J_5' = \partial^4 + 10(u_{xx} + u\partial^2 + u_x\partial + 3u^2),\ J_5'[1] = 10J_3,\ J_5' - J_5'[1] = \partial^4 + 10\partial u\partial$$

**5.** Here we present the properties of $X^n = X^n_{(+)} + X^n_{(-)}$, where $X^n_{(+)}$ is differential and $X^n_{(-)}$ integral part (the terms with $\partial^{-1}$) of $X^n$, $n = 1, 2, \ldots$. A few examples

$\quad X_{(+)} = \partial \qquad\qquad\qquad X_{(-)} = u\partial^{-1}$

$\quad X^2_{(+)} = \partial^2 + 2u \qquad\quad X^2_{(-)} = u_x\partial^{-1} + u\partial^{-1}u\partial^{-1}$

$\quad X^3_{(+)} = \partial^3 + 3\partial u \qquad X^3_{(-)} = (u_{xx} + 2u^2)\partial^{-1} + u\partial^{-1}u + u_x\partial^{-1}u\partial^{-1} + u\partial^{-1}u\partial^{-1}u\partial^{-1}$

$\quad X^4_{(+)} = \partial(\partial^3 + 3\partial u) + \partial^2 u + 6u^2 - 2u_x\partial$

$\quad X^5_{(+)} = \partial^5 + 5\partial(\partial^2 u - u_x\partial + 2u^2)$

As is seen, $X_{(+)}$, $X^3_{(+)}$ and $X^5_{(+)}$ are $\partial O$. $X^n_{(-)}$ can be expressed via $f_n$ by using integration by parts: $\partial^{-1}u\partial^{-1} = p\partial^{-1} - \partial^{-1}p$ leads to $X^2_{(-)} = f_2\partial^{-1} - u\partial^{-1}p$; similarly we have $X^3_{(-)} = f_3\partial^{-1} - f_2\partial^{-1}p + u\partial^{-1}\varphi_2$. These examples illustrate the following properties of $X^n_{(+)}$ and $X^n_{(-)}$ (proven in Appendix):

**P1**. $X^m_{(+)} = \partial P_m$, $P_m$ is a self-adjoint differential operator: $uP_m - (P_mu) = \partial O_m$

**P2**. (a) $X^m_{(+)}u = K_m$, so $(P_mu) = J_m$ \qquad (b) $X^{m+1}_{(+)} = \partial X^m_{(+)} + uP_m + J_m$

**P3**. $[X^m_{(+)}, X] = K_m\partial^{-1}$

**P4**. $X^n_{(-)} = f_n\partial^{-1} - f_{n-1}\partial^{-1}\varphi_1 + f_{n-2}\partial^{-1}\varphi_2 - \ldots + f_1\partial^{-1}\varphi_{n-1}$

(Some of them are considered in [4]). These properties lead to the following results.

- Applying the commutator $[X^m_{(+)}, X] = K_m\partial^{-1}$ from P3 to $f_n$ gives the equations, for each m,
$$X^m_{(+)}f_{n+1} = XX^m_{(+)}f_n + K_m\varphi_n \qquad (18)$$
Starting from $X^m_{(+)}f_1 = K_m$, by P2(a), their r.h.s's can be recursively expressed through $K_m$ for all n. The first two of them are

$\quad X^m_{(+)}f_2 = XK_m + K_mp \equiv [K_m + p(\partial^{-1}K_m)]_x$

$\quad X^m_{(+)}f_3 = XX^m_{(+)}f_2 + K_m\varphi_2 \equiv 2uK_m + [X^m_{(+)}f_2 + (p^2/2)(\partial^{-1}K_m)]_x$

These r.h.s's are $f_2'[K_m]$ and $f_3'[K_m]$, respectively. In general, for all n, and each m
$$X^m_{(+)}f_n = f_n'[K_m] \qquad (19)$$
Proof. By induction. If this is true, then (18) becomes $f_{n+1}'[K_m] = Xf_n'[K_m] + \varphi_nK_m$. This is also true because always $(Xf)' = Xf' + (\partial^{-1}f)$.

Adding $(f_n)_t$ to both sides of (19) gives $(f_n)_t + X^m_{(+)}f_n = f_n'[U_m]$. With $U_m = 0$, it yields
$$(f_n)_t + X^m_{(+)}f_n = 0$$
Since $X^m_{(+)}f_n = ED$, due to P1, these equations represent the other form of the conservation laws (17b) and trivial ones with n even.

- Using (19) and P1 we can prove differently (cf. §4) that for each m and any k
$$J_k'[K_m] = ED$$
Proof. We write J for convenience as $J = \Phi_o + 2\Phi_o\varphi_e - (\varphi_o\varphi_e)_x$. Then
$$J'[K_m] = \Phi_o'[K_m] + 2\varphi_e\Phi_o'[K_m] + 2\Phi_o\varphi_e'[K_m] - ((\varphi_o\varphi_e)'[K_m])_x$$
The last term is ED explicitly. $\Phi_o'[K_m] = (P_m\Phi_o)_x$ and $\varphi_e'[K_m] = P_m\Phi_e$ by (19). With that, the second term can be written as $2[(\varphi_eP_m\Phi_o)_x - \Phi_eP_m\Phi_o]$. Then the r.h.s is ED plus $2(\Phi_oP_m\Phi_e - \Phi_eP_m\Phi_o)$, which is also ED because $P_m$ is self-adjoint.

- From the above properties of $X^n$ it also follows the recurrence (see Appendix)
$$P_{m+2} = \partial^2 P_m + uP_m + J_m + \partial^{-1}(u\partial P_m + uO_m + K_m) \qquad (20)$$
Since $(P_mu) = J_m$, by P2(a), and $(O_mu) = 0$, then $(P_{m+2}u) = (\partial^2 + 2u + 2\partial^{-1}u\partial)(P_mu)$, which is $J_{m+2} = RJ_m$. This result represents the *definition* of recursion operator R in the framework of the present approach.

- The relation (20) also leads to the recursive formula (21) for densities as follows.

Applying $uP_m - J_m = \partial O_m$, see P1, 2, to $f = 1$ gives $u(P_m 1) = J_m + (O_m 1)_x$. Since $J_k = k\rho_k + ED$, by (15), then $u(P_m 1) = m\rho_m + ED$ (m = k here). Examples

$(P_3 1) = (\partial^2 + 3u)1 = 3u$, $u(P_3 1) = 3\rho_3$

$(P_5 1) = [\partial^4 + 5(\partial^2 u - u_x \partial + 2u^2)]1 = 5(u_{xx} + 2u^2)$, $u(P_5 1) = 5\rho_{5D}$

The next $u(P_m 1)$ are not such simple. To find them we do not need to know $P_{m+2}$ in (20) in full (namely $O_m$), we can use the recursive formula following from it

$$(P_{m+2} 1) = (P_m 1)_{xx} + u(P_m 1) + 2J_m + \partial^{-1} u[(P_m 1)_x + (O_m 1)] \tag{21}$$

where $(O_m 1)$ is to be found from $(O_m 1)_x = u(P_m 1) - J_m$. Examples

With $(P_5 1)$ known we get $(O_5 1)_x = u(P_5 1) - J_5 = -[u_{xx} + (5/2)u^2]_{xx}$ and then (21) yields

$(P_7 1) = 7(u^{(4)} + 5u^3 + 7uu_{xx} + 4u_x^2)$      $u(P_7 1) = 7\rho_7 + 7[uu_{xx} - u_x^2 + (7/3)u^3]_{xx}$

Next: $(O_7 1)_x = u(P_7 1) - J_7 = -[u^{(4)} + (7/2)(u^2)_{xx} + 7u^3]_{xx}$ leads to

$u(P_9 1) = 9u(u^{(6)} + 14u^4 - 70uu_x^2 + 10uu^{(4)} - 6u_{xx}^2) + [336u^3 u_x + 204uu_x u_{xx} - 68u_x^3]_x$

Thus found density is equivalent to $\rho_9$. And, of course, $u(P_m 1)$ and $G_m$ (§3) are equivalent.

**Highlights**

- The fluxes are given by $J = \Phi_o + \Phi_o \varphi_e - \Phi_e \varphi_o$ and $J_m = (P_m u)$, $P_m \equiv \partial^{-1} X^m{}_{(+)}$.
- The densities $\rho_{kD}$ are given by $f_k = \rho_{kD} + ED$ (and equivalent to the known densities $\rho_k$).
- The term $k\rho_{kD}$ is a nonderivative part of $J_k$ and $u(P_k 1)$.
- The conservation laws of each equation $U_m = 0$ have the form $g_t + g'[K_m] = 0$, $g = \rho_{kD}$, $f_k$, $J_k$; wherein $f_k'[K_m] = X^m{}_{(+)} f_k$.
- The recursion operator R for $J_m$ appears from relation between adjacent operators $P_m$.

**Appendix**

To express and prove the properties of $X^n{}_{(+)}$ and $X^n{}_{(-)}$ formulated in §5, it is convenient to introduce the operators $F_{o,e}$, such that $F_{o,e} u = \Phi_{o,e}$. They are the series $\zeta F_o = \Sigma z^i X^i$, $i = 0, 2, 4, \ldots$, and $\zeta F_e = \Sigma z^m X^m$, $\zeta \equiv 1/z$. They are related as (cf. (6))

$$XF_o = \zeta F_e \qquad XF_e = \zeta F_o - 1 \tag{A1}$$

Then P4 can be written as

$$\zeta F_{e(-)} = \Phi_o \partial^{-1} + \Phi_o \partial^{-1} \varphi_e - \Phi_e \partial^{-1} \varphi_o = \Phi_o \partial^{-1} + \partial(\varphi_o \partial^{-1} \varphi_e - \varphi_e \partial^{-1} \varphi_o) \tag{A2}$$

$$\zeta F_{o(-)} = \Phi_e \partial^{-1} + \Phi_e \partial^{-1} \varphi_e - \Phi_o \partial^{-1} \varphi_o \tag{A3}$$

(i) First we prove the property P1. We show that the operator $B \equiv \partial^{-1} F_e$ (the series of $\partial^{-1} X^m$) is self-adjoint. Then, clearly, its parts $B_{(+)}$ and $B_{(-)}$ are such *separately*.

Proof. We use the easily verified formulas

$$X^{2\dagger} = \partial^{-1} X^2 \partial \qquad X^\dagger \partial^{-1} = -\partial^{-1} X \tag{A4}$$

where $X^{2\dagger} = \partial^2 + 2u - \partial^{-1} u_x + \partial^{-1} u \partial^{-1} u$, $X^\dagger = -\partial - \partial^{-1} u$; and obvious $X^n F_{e,o} = F_{e,o} X^n$.

From (A1): $X^2 F_e = F_e X^2 = \zeta^2 F_e - X$. Integration of this gives $\partial^{-1} X^2 \partial B = BX^2 = \zeta^2 B - \partial^{-1} X$. Adjoint of the second equality is $X^{2\dagger} B^\dagger = \zeta^2 B^\dagger + X^\dagger \partial^{-1}$. This, by (A4), can be put as $\partial^{-1} X^2 \partial B^\dagger = \zeta^2 B^\dagger - \partial^{-1} X$. Comparison with the initial form yields the proof: $B^\dagger = B$.

Hence the series $\zeta \partial^{-1} F_{e(+)} \equiv P = \Sigma z^m P_m$ and thereby each its member $\partial^{-1} X^m{}_{(+)} \equiv P_m$ is self-adjoint. $\{\partial^{-1} X^3{}_{(+)} \equiv P_3 = \partial^2 + 3u\}$. For P, we express this fact as

$$uP - (Pu) = \partial O_P \tag{A5}$$

where $O_P = \Sigma z^m O_m$ with operators $O_m$ defined in P1. Also, from (A2) (proven below) it follows that $\zeta B_{(-)} \equiv \zeta \partial^{-1} F_{e(-)} = \partial^{-1} \Phi_o \partial^{-1} + \varphi_o \partial^{-1} \varphi_e - \varphi_e \partial^{-1} \varphi_o$, which is self-adjoint as well.

(ii) Here we assume that $F_{e(-)}$ in (A2) is true.
First we find $F_{o(-/+)}$ from $\zeta F_o = XF_{e(+)} + XF_{e(-)} + 1$, see (A1). Since $\zeta F_{e(+)} = \partial P$ then
$$\zeta X F_{e(+)} = \partial^2 P + uP \tag{A6}$$
Using (A2), the formulas $J \equiv \Phi_o + \Phi_o \varphi_e - \Phi_e \varphi_o$ and $\partial^{-1} f_x \partial^{-1} = f \partial^{-1} - \partial^{-1} f$, we obtain
$$\zeta X F_{e(-)} = J + (X\Phi_o)\partial^{-1} + (X\Phi_o)\partial^{-1}\varphi_e - (X\Phi_e + u)\partial^{-1}\varphi_o$$
This, due to $X\Phi_o = \zeta \Phi_e$ and $X\Phi_e = \zeta \Phi_o - u$, see (6), is
$$\zeta X F_{e(-)} = J + \zeta(\Phi_e \partial^{-1} + \Phi_e \partial^{-1} \varphi_e - \Phi_o \partial^{-1} \varphi_o) \tag{A7}$$
As a result, $\zeta F_{o(-)} = (XF_{e(-)})_{(-)}$ is the same as in (A3). In addition we have the proof of P2(b)
$$\zeta F_{o(+)} = z(\partial^2 P + uP + J) + 1, \text{ where } 1 = \zeta F_o(z^0) \tag{A8}$$
Second, using (A2) and taking $u\varphi_{e,o}$ from (6), we have
$$\zeta F_{e(-)} u = \Phi_o p + \Phi_o \partial^{-1}(\zeta \Phi_o - \Phi_{ex} - u) - \Phi_e \partial^{-1}(\zeta \Phi_e - \Phi_{ox}) = \zeta(\Phi_o \varphi_o - \Phi_e \varphi_e) = (\zeta/2)(\varphi_o^2 - \varphi_e^2)_x$$
Therefore, from $\Phi_e \equiv F_{e(+)} u + F_{e(-)} u = [zJ + (1/2)(\varphi_o^2 - \varphi_e^2)]_x$, see (7), we have the proof of P2(a)
$$\zeta F_{e(+)} u = K \quad \text{hence} \quad (Pu) = J \tag{A9}$$
The proofs (A8) and (A9) are based on the assumption about (A2) and hence are *pending*. As we will see, only if they are true the noncontradictory, self-consistent results can be obtained.

(iii) Now we derive $F_{e(-)}$ in (A2) back from $F_{o(-)}$ in (A3) and thereby prove P4.
From (A1): $\zeta F_e = XF_{o(+)} + XF_{o(-)}$. Here, due to (A3) and then (6),
$$\zeta X F_{o(-)} = zK + (\zeta\Phi_o - u)\partial^{-1} + (\zeta\Phi_o - u)\partial^{-1}\varphi_e - \zeta\Phi_e\partial^{-1}\varphi_o - u\partial^{-1}(\varphi_e + \varphi_e^2 - \varphi_o^2)$$
The use of $2\varphi_e = z2J + (\varphi_o^2 - \varphi_e^2)$, due to (7), yields
$$\zeta X F_{o(-)} = zK + \zeta(\Phi_o\partial^{-1} + \Phi_o\partial^{-1}\varphi_e - \Phi_e\partial^{-1}\varphi_o) - zu\partial^{-1}2J - u\partial^{-1}$$
By (A8): $\zeta^2 X F_{o(+)} = \partial(\partial^2 P + uP + J + \zeta) + u\partial P + u\partial^{-1}(uP + J + \zeta)$. As is seen,
$\zeta^2(XF_{o(+)})_{(-)} = u\partial^{-1}(J + \zeta) + u(\partial^{-1}uP)_{(-)}$, where $uP = (Pu) + \partial O_P$, by (A5). Only if $(Pu) = J$, cf. (A9), then $\zeta(XF_{o(+)})_{(-)} = zu\partial^{-1}2J + u\partial^{-1}$, which cancels in $\zeta(XF_o)_{(-)}$ giving self-consistently $\zeta F_{e(-)}$ in (A2). Now when (A2) is proved, (A8) is justified. ($F_{e(+)}$ following from these formulas is given in (A15)).

(iv) Here we find commutator $[F_{e(+)}, X]$. From the obvious equality $[X, F_e] = 0$ we have
$$[F_{e(+)}, X] = [X, F_{e(-)}] \quad \text{and hence} \quad [F_{e(+)}, X]_{(+/-)} = [X, F_{e(-)}]_{(+/-)} \tag{A10}$$
By (A6): $\zeta[F_{e(+)}, X] = \partial P\partial + \partial Pu\partial^{-1} - (\partial^2 P + uP)$. To find $(\partial Pu\partial^{-1})_{(+/-)}$, we use the equality
$$Pu - (Pu) = Z\partial, \text{ where Z is an operator,} \tag{A11}$$
valid for *any* operator in place of P - such form ensures the identity when (A11) applied to $f = 1$. Owing to this, $Pu\partial^{-1} = (Pu)\partial^{-1} + Z$ and $\partial Pu\partial^{-1} = (Pu)_x \partial^{-1} + (Pu) + \partial Z$. Thus
$$\zeta[F_{e(+)}, X]_{(-)} = (Pu)_x \partial^{-1} \qquad \zeta[F_{e(+)}, X]_{(+)} = \partial(P\partial - \partial P + Z) - uP + (Pu) \tag{A12a,b}$$
Now we find $[X, F_{e(-)}]$. By (A2)
$$\zeta F_{e(-)} X = \Phi_o + \Phi_o\partial^{-1}\varphi_e\partial - \Phi_e\partial^{-1}\varphi_o\partial + \zeta F_{e(-)} u\partial^{-1} = J - \Phi_o\partial^{-1}\Phi_e + \Phi_e\partial^{-1}\Phi_o + \zeta F_{e(-)}u\partial^{-1}$$
This and $\zeta X F_{e(-)}$ from (A7) yield
$$\zeta[X, F_{e(-)}] = \zeta(\Phi_e\partial^{-1} + \Phi_e\partial^{-1}\varphi_e - \Phi_o\partial^{-1}\varphi_o) - (\Phi_e\partial^{-1}\Phi_o - \Phi_o\partial^{-1}\Phi_e) - \zeta F_{e(-)}u\partial^{-1}$$
Integration by parts $\partial^{-1} f = f\partial^{-1} - \partial^{-1} f_x \partial^{-1}$ in all but the first and last terms, and formulas (A2) and (6) gives $\zeta[X, F_{e(-)}] = \zeta(\Phi_e + \Phi_e\varphi_e - \Phi_o\varphi_o)\partial^{-1} \equiv K\partial^{-1}$. Then (A10), (A12a) and P1, $\zeta F_{e(+)} = \partial P$, lead to the proof of P2(a) (and justify (A9))
$$(Pu)_x = \zeta F_{e(+)} u = K \tag{A13}$$
As is seen, $\zeta[X, F_{e(-)}]_{(+)} = 0$, hence $\zeta[F_{e(+)}, X]_{(+)} = 0$ too. This and (A13) provide the proof of P3

$\zeta[F_{e(+)}, X] = \zeta[F_{e(+)}, X]_{(-)} = K\partial^{-1}$

Since $\zeta[F_{e(+)}, X]_{(+)} = 0$ then (A12b) gives $uP - (Pu) = \partial(P\partial - \partial P + Z)$. This means that $P = P^{\dagger}$, and, due to (A5), that

$$O_P = P\partial - \partial P + Z \qquad (A14)$$

(v) Here we show that $F_{e(+)}$ following from the block (iii)

$$\zeta^3 F_{e(+)} = \zeta^2(XF_{o(+)} + XF_{o(-)})_{(+)} = \partial(\partial^2 P + uP + J + \zeta) + u\partial P + (u\partial^{-1}uP)_{(+)} + K \qquad (A15)$$

is $\partial O$ and thereby prove the property P1, in essence, by induction. For this, it is suffice to show that the last three terms, denote them as W, form $\partial O$, with O being a differential operator.

By (A5): $\partial^{-1}uP = O_P + \partial^{-1}(Pu)$, so $(u\partial^{-1}uP)_{(+)} = uO_P = u(P\partial - \partial P + Z)$, by (A14). So,
$W = u(P\partial + Z) + K = [\partial O_P + (Pu)]\partial + uZ + K = \partial O_P\partial + \partial J + uZ$, as $(Pu) = J$, by (A13). Thus, it remains to prove that $uZ$ is $\partial O$. Comparing the adjoint of (A5), $Pu - (Pu) = -O_P^{\dagger}\partial$, and (A11) gives $Z = -O_P^{\dagger}$. Using this and $(O_Pu) = 0$ in the general equality $uZ - (Z^{\dagger}u) = \partial O_Z$, with an operator $O_Z$, we obtain $uZ = \partial O_Z$. As a result, (A15) acquires the sought-out form

$$\zeta^3 F_{e(+)} = \partial(\partial^2 P + uP + 2J + O_P\partial + O_Z + \zeta)$$

Equation (A15) for individual members, with $(u\partial^{-1}uP_m)_{(+)} = uO_m$, reads

$$X^{m+2}_{(+)} = \partial(\partial^2 P_m + uP_m + J_m) + u\partial P_m + uO_m + K_m$$

Integration of this gives the recurrence (20).

Finding $W_3$ and $uZ_3$: $P_3 = \partial^2 + 3u \rightarrow uP_3 - J_3 = u\partial^2 - u_{xx} = \partial O_3$, $O_3 = u\partial - u_x \rightarrow$
$W_3 \equiv u\partial P_3 + uO_3 + K_3 = u[\partial(\partial^2 + 3u) + u\partial - u_x] + u_{xxx} + 3(u^2)_x = \partial^3 u - 3\partial u_x\partial + 4\partial u^2$.
$Z_3 = -O_3^{\dagger} = \partial u + u_x = u\partial + 2u_x \rightarrow uZ_3 = \partial u^2$.

E-mail: bryssev@gmail.com